\input lanlmac
\input epsf.tex
\input mssymb.tex
\overfullrule=0pt

\newcount\figno
\figno=0
\def\fig#1#2#3{
\par\begingroup\parindent=0pt\leftskip=1cm\rightskip=1cm\parindent=0pt
\baselineskip=11pt
\global\advance\figno by 1
\midinsert
\epsfxsize=#3
\centerline{\epsfbox{#2}}
\vskip 12pt
{\ninebf Fig.\ \the\figno:} {\ninerm #1}\par
\endinsert\endgroup\par
}
\def\figlabel#1{\xdef#1{\the\figno}%
\writedef{#1\leftbracket \the\figno}%
}
\def\omit#1{}

\def\e#1{{\rm e}^{#1}}
\def\pre#1{{\tt
#1}}%use this to give preprint # in refs
\def\tr{{\rm tr}}
\def\Rc{{\check R}}

\def\oh{{1\over 2}}

\def\ket#1{\left| #1 \right>}
\def\braket#1#2{\left< #1 | #2 \right>}
\def\CN{{\cal N}}
\def\Y{{Y_{n+1}}}\def\pY{{Y'_n}}\def\ppY{{Y_{n}}}

\def\frown{{\hbox{\epsfbox{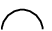}}}}
\def\lharch{\rceil}\def\rharch{\lceil}
\def\rharch{{\hbox{\epsfbox{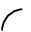}}}}
\def\lharch{{\hbox{\epsfbox{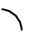}}}}
\def\sqt{\sqrt{3}}\def\sqs{\sqrt{7}}
\def\Arctan{{\rm Arctan\,}}

\def\qed{\nobreak\hfill\vbox{\hrule height.4pt%
\hbox{\vrule width.4pt height3pt \kern3pt\vrule width.4pt}\hrule height.4pt}\medskip\goodbreak}
\lref\AS{F.~Alcaraz and Y.~Stroganov, {\it The Wave Functions for the Free-Fermion 
Part of the Spectrum
of the $SU_q(N)$ Quantum Spin Models}, \pre{cond-mat/0212475}.}
\lref\BdGNun{M.T. Batchelor, J. de Gier and B. Nienhuis,
{\sl The quantum symmetric XXZ chain at $\Delta=-1/2$, alternating sign matrices and 
plane partitions},
{ J. Phys.} A34 (2001) L265--L270,
\pre{cond-mat/0101385}.}
\lref\Bre{D. Bressoud, {\sl Proofs and Confirmations: The Story of
the Alternating Sign Matrix Conjecture}, Cambridge Univ. Pr., 1999.}
\lref\DF{P.~Di Francesco, {\sl 
 A refined Razumov--Stroganov conjecture} I: 
     J. Stat. Mech. P08009 (2004), \pre{cond-mat/0407477}; II: 
%PDF  A refined Razumov--Stroganov conjecture: II 
     J. Stat. Mech. P11004 (2004), \pre{cond-mat/0409576}.}
\lref\DFa{P.~Di Francesco, {\sl 
 Inhomogeneous loop models with open boundaries }, { J. Phys. A: Math. Gen.} {\bf 38} 6091 (2005), \pre{math-ph/0504032}}
%\semi
\lref\DFb{P.~Di Francesco, 
{\sl Boundary $q$KZ equation and generalized Razumov--Stroganov sum rules for open IRF models}, J. Stat. Mech. P11003 (2005), \pre{math-ph/0509011}.}
\lref\DFKZJ{P.~Di Francesco, A. Knutson and P. Zinn-Justin, work in progress.}
\lref\DFZJ{P.~Di Francesco and P.~Zinn-Justin, {\sl Around the Razumov--Stroganov conjecture:
proof of a multi-parameter sum rule}, { E. J. Combi.} 12 (1) (2005), R6,
\pre{math-ph/0410061}.}
\lref\DFZJb{P.~Di Francesco and P.~Zinn-Justin, {\sl Inhomogeneous model of crossing loops
and multidegrees of some algebraic varieties}, 
%Commun. Math. Phys. 262 (2006), 459--487
{Commun. Math. Phys}  {\bf 262}  459--487, (2006), %\break
\pre{math-ph/0412031}.}
\lref\DFZJc{P.~Di Francesco and P.~Zinn-Justin, {\sl The 
quantum Knizhnik--Zamolodchikov equation, 
generalized Razumov--Stroganov sum rules and extended Joseph polynomials }, 
{J. Phys. A } {\bf 38} %No 48 (2 December 2005)
L815-L822  (2006), \pre{math-ph/0508059}.}
\lref\DFZJd{P.~Di Francesco and P.~Zinn-Justin, {\sl From Orbital Varieties to Alternating 
Sign Matrices}, to appear in the proceedings of FPSAC'06 (2006), \pre{math-ph/0512047}.}
\lref\dG{J. de Gier, {\sl The art of number guessing: where
combinatorics meets physics}, preprint\pre{math.CO/0211285}.}
\lref\dGN{J. de Gier and B. Nienhuis, {\sl Brauer loops and 
the commuting variety}, \pre{math.AG/0410392}. }
\lref\FR{I.B.~Frenkel and N.~Reshetikhin, {\sl Quantum affine Algebras and Holonomic 
Difference Equations},
{Commun. Math. Phys.} 146 (1992), 1--60.}
\lref\Ho{R.~Hotta, {\sl On Joseph's construction of Weyl group 
representations}, Tohoku Math. J. Vol. 36 (1984), 49--74.}
\lref\JM{M.~Jimbo and T.~Miwa, {\it Algebraic analysis of Solvable Lattice Models}, 
CBMS Regional Conference Series in Mathematics vol. 85, American Mathematical Society, 
Providence, 1995.}
\lref\Jo{A. Joseph, {\sl On the variety of a highest weight module}, 
{ J. Algebra} 88 (1) (1984), 238--278.}
\lref\KZ{V.~Knizhnik and A. Zamolodchikov, {\sl Current algebra and Wess--Zumino 
model in two dimensions},
{Nucl. Phys.} B247 (1984), 83--103.}
\lref\KZJ{A. Knutson and P. Zinn-Justin, {\sl A scheme related to the Brauer loop model}, 
\pre{math.AG/0503224}.}
\lref\Ku{G. Kuperberg, {\sl Symmetry classes of alternating-sign
matrices under one roof}, 
{ Ann. of Math.} 156 (3) (2002), 835--866,
\pre{math.CO/0008184}.}
\lref\Lu{G. Luzstig, {\sl Canonical bases arising from quantized enveloping algebras}, 
{J. Amer. Math. Soc.} 3 (1990), 447--498.}
\lref\Mitretal{S. Mitra, B. Nienhuis, J. de Gier, M. T. Batchelor, 
{\sl Exact expressions for correlations in the ground state of the dense 
O(1) loop model}, JSTAT (2004) P09010,  
\pre{cond-mat/0401245}}.
\lref\Oka{S. Okada, {\sl  Enumeration of Symmetry Classes of Alternating Sign Matrices and Characters of Classical Groups}, \pre{math.CO/0408234}.}
\lref\Pas{V.~Pasquier, 
{\sl Quantum incompressibility and Razumov Stroganov type conjectures},% <<<
\pre{cond-mat/0506075};
{\sl Incompressible representations of the Birman-Wenzl-Murakami algebra},
\pre{math.QA/0507364}.}
\lref\PRdG{ P. A. Pearce, V. Rittenberg and J. de Gier,
{\sl Critical Q=1 Potts Model and Temperley--Lieb Stochastic Processes},
\pre{cond-mat/0108051 }.}
\lref\PRdGN{ P. A. Pearce, V. Rittenberg, J. de Gier and B. Nienhuis,
{\sl Temperley--Lieb Stochastic Processes}, 
{ J. Phys. A} {\bf 35 } (2002) L661-L668, \pre{math-ph/0209017}.}
\lref\Ro{W. Rossmann,
{\sl Equivariant multiplicities on complex varieties.
  Orbites unipotentes et repr\'esentations, III},
  { Ast\'erisque} No. 173--174, (1989), 11, 313--330.}
\lref\Rob{D.P. Robbins, {\sl Symmetry classes of Alternating Sign Matrices},
\pre{math.CO/0008184}.}
\lref\RSun{A.V. Razumov and Yu.G. Stroganov, 
{\sl Spin chains and combinatorics}, 
J. Phys. A {\bf 34} (2001) 3185-3190,
\pre{cond-mat/0012141}}
\lref\RSunb{A.V. Razumov and Yu.G. Stroganov, 
{\sl Spin chains and combinatorics: twisted boundary conditions}, 
J. Phys. A {\bf 34} (2001) 5335-5340, 
\pre{cond-mat/0102247}.}
\lref\RSde{A.V. Razumov and Yu.G. Stroganov, 
{\sl Combinatorial nature
of ground state vector of $O(1)$ loop model},
{Theor. Math. Phys.} 
{\bf 138} (2004) 333--337; { Teor. Mat. Fiz.} 138 (2004) 395--400, 
\pre{math.CO/0104216}.}
\lref\RStr{A.V. Razumov and Yu.G. Stroganov, 
{\sl $O(1)$ loop model with different boundary conditions and symmetry classes of alternating-sign matrices},
{Theor. Math. Phys.} 
{\bf 142} (2005) 237--243; {Teor. Mat. Fiz.} 142 (2005) 284--292,
\pre{cond-mat/0108103}.}
\lref\RSqu{A.V. Razumov and Yu.G. Stroganov, 
{\sl Enumerations of half-turn symmetric alternating-sign matrices 
of odd order}, \pre{math-ph/0504022}.}

\Title{SPhT-T06/020}
{\vbox{
\centerline{Sum rules for the ground states}
\medskip
\centerline{of the $O(1)$ loop model on a cylinder}
\medskip
\centerline{and the XXZ spin chain}
}}
\bigskip\bigskip
\centerline{P.~Di~Francesco \footnote{${}^\#$} 
{Service de Physique Th\'eorique de Saclay,
CEA/DSM/SPhT, URA 2306 du CNRS,
C.E.A.-Saclay, F-91191 Gif sur Yvette Cedex, France, 
{\tt Philippe.Di-Francesco@cea.fr}}, 
P.~Zinn-Justin \footnote{${}^\star$}
{Laboratoire de Physique Th\'eorique et Mod\`eles Statistiques, UMR 8626 du CNRS,
Universit\'e Paris-Sud, B\^atiment 100,  F-91405 Orsay Cedex, France, 
{\tt pzinn@lptms.u-psud.fr}}
and J.-B.~Zuber \footnote{${}^{\bullet\,\#}$} % <<<
{Laboratoire de Physique Th\'eorique et Hautes \'Energies, Tour 24-25 
5\`eme \'etage,
Universit\'e Pierre et Marie Curie-Paris6; UMR 7589 du CNRS; Universit\'e 
Denis Diderot-Paris7, Bo\^\i te 126,
4 pl Jussieu, 75252 Paris Cedex 05, France, {\tt zuber@lpthe.jussieu.fr}} 
}
%\bigskip
\vskip0.5cm
%abstract

\noindent The sums of components of the ground states
of the $O(1)$ loop model on a cylinder or of the XXZ 
quantum spin chain at $\Delta=-\oh$ of size $L$ are
expressed in terms of 
combinatorial numbers. The methods include
the introduction of spectral parameters and the use of 
integrability, a mapping from size $L$ to $L+1$, 
and knot-theoretic skein relations.

\bigskip

\def\today{\number\day\ \ifcase\month\or January \or February \or March \or
April \or May \or June \or July \or August \or September \or October \or
November \or December\fi\space\number\year}   
%AMS Subject Classification (2000): Primary 05A19; Secondary 82B20
%\draft
\Date{03/2006}
%\Date{\today}
%
%%%%%%%%%%%%%%%%%%%%%%%%%%%%%%%%%%%%%%%%%%%%%%%%%%%%%%%%%%%%%%%%%%%%%
%
%\input intro.tex

\newsec{Introduction and summary of results}
\par\noindent
The observation by Razumov and Stroganov \RSun\  that the components of the 
ground state of integrable quantum spin chains in some adequate basis
enjoy integrality conditions and are connected to known combinatorial 
problems, to wit, alternating sign matrices (ASM) and their avatars, 
has been the source of an amazing burst of new developments. 
It has been soon realized that in many cases
these integers in fact count some configurations of other
lattice models \refs{\BdGNun, \RSde}. To be more explicit, the components 
of the ground state of the Temperley--Lieb Hamiltonian describing 
the $O(1)$ loop model 
on an even number of sites with periodic boundary conditions
are conjectured to count the numbers of configurations of so-called
fully packed loop (FPL) models: this is the now celebrated 
Razumov--Stroganov conjecture \RSde.
Other types of boundary conditions have also been considered 
\refs{\PRdG,\RStr,\PRdGN}.

 After these original observations and conjectures, 
a major progress has been the introduction of inhomogeneities
in the original problem, in the form of spectral parameters, thus enabling 
one to use the full machinery of integrable models~\refs{\DF,\DFZJ}. In 
particular, this has led to recursion formulae between components of 
the ground state and ultimately to a proof of the 
ground state sum rule, a weak but non trivial 
version of the RS conjecture. These recursion formulae have been 
shown to follow from an underlying algebraic structure rooted in the 
Affine Hecke Algebra (AHA) \Pas, or alternatively, related to
the so-called quantum Knizhnik--Zamolodchikov equation (\DFZJc\ and further 
references therein). % [DFZJ-0508059].
This has then been extended into several distinct directions. 
Loop models with crossings \dGN\ have been shown to display an amazing 
relationship to the algebraic geometry of matrix varieties \refs{\DFZJb,\KZJ}. 
Extensions to other types of boundary conditions~\DFa, %[open DF-0504032],
to higher rank  algebras \DFZJc, or to both \DFb, have been considered in turn.

In the present paper, we return to a case first tackled in 
the original papers in this domain, that of   
the $O(1)$ loop model on a square lattice wrapped on a 
semi-infinite cylinder of odd integer perimeter 
\refs{\BdGNun,\RStr}%[BadGN 0101385]
, or of the associated periodic XXZ spin chain \RSun. %[RS0012141].
In the loop model, the sites of the boundary of the cylinder
are pairwise connected via non-intersecting links. 
The oddness of the perimeter implies 
that one site remains unmatched, hence giving rise to a
defect line, connecting it to the point at infinity 
on the cylinder. In the spin chain on $L=2n+1$ sites, likewise, 
 the ground state 
is made of $n+1$ spins pointing upward and $n$ downward (or vice-versa).
It is also interesting to consider the case of the $O(1)$ loop model
on a cylinder of even perimeter, when the loops wrapping around the cylinder
are not allowed to contract, 
or equivalently, when the corresponding link pattern is drawn on a
punctured  disk. This case was called periodic with ``distinct connectivities''
in \PRdGN. It must be distinguished from the more usual
loop model with ``identified connectivities'' which was the subject
of \DFZJ, which will be also discussed in what follows since we need it
to analyze the other cases. The related XXZ
spin chain of even size $L=2n$ has twisted boundary conditions \BdGNun,
as we recall below in sect. 4. See also \Mitretal\ for more data 
on these different boundary conditions.

Let $A_n=\prod_{j=1}^n{(3j-2)!\over(n+j-1)!}$ denote the number 
of alternating sign matrices of size $n$ (\Bre\ and further references 
therein) and $\CN_n$ stand for 
\eqn\calNn{
\CN_n={3^{n/2}\over 2^n} {2\times5\times\cdots\times(3n-1)\over 1\times3\cdots\times(2n-1)} A_n
=3^{n/2}{\prod_{j=1}^{n}(3j-1)!\over\prod_{j=1}^n (n+j)!}\ .}
Let us also introduce the number $A_{{\rm HT}}(L)$ of ``half-turn 
symmetric''  alternating sign matrices \refs{\Bre,\Rob,\Ku,\RSqu}
\eqn\AHT{A_{{\rm HT}}(L)= 
\cases{\prod_{j=0}^{n-1}{3j+2\over 3j+1}\left({(3j+1))!\over (n+j)!}
\right)^2%=2,10,140,\cdots 
 & if $L=2n$ is even\cr
\cr
\prod_{j=1}^{n}{4\over 3} \left({(3j)!j!\over(2j)!^2}\right)^2
%=1,3,25,\cdots
 & if $L=2n+1$ is odd\ .\cr
}}
Okada \Oka\ showed that this number $ A_{{\rm HT}}(L)$ may 
be expressed in terms of the dimension of certain representations of 
GL$(L)$
\eqn\formOk{\eqalign{
 A_{{\rm HT}}(2n+1)&=3^{-n^2} (\dim_{\pY}^{{\rm GL}(2n+1)})^2\cr
 A_{{\rm HT}}(2n)&=3^{-n(n-1)} 
\dim_{\ppY}^{{\rm GL}(2n)} 
\dim_{\pY}^{{\rm GL}(2n)}
}\ .}
The Young diagram $\pY$ which defines that representation is made of
one row of length $n$, two rows of length $n-1$, two of length $n-2$, \dots, 
two of length 1, while the Young diagram $\ppY$ results from the 
ablation of the first row of $\pY$.
%\foot{Okada actually expressed 
%$A_{{\rm HT}}(2n)$ in terms of $Y'$ and another representation, whose 
%dimension is the same as $Y''$ 
%We find it more convenient to use $Y''$ in view of the
%following developments. }

In \BdGNun, %[BadGN 0101385], 
it was conjectured 
that for the $O(1)$  loop model on a cylinder of perimeter $L=2n+1$, 
all components of the Perron--Frobenius eigenvector $\Psi$ are integers 
if the smallest one is normalized to be 1, that 
the sum of components of $\Psi$ over all loop configurations $\pi$ is
\eqn\sumloop{
\sum_\pi \Psi_\pi = A_{{\rm HT}}(2n+1)=
\CN_n^2=1,3,25,588,39204\ldots\qquad L=2n+1=1,3,\ldots}
and that the largest component is
\eqn\psimloop{\Psi_{max}=A_n^2=1,1,4,49,1764, \ldots\qquad L=2n+1=1,3,\ldots}

As for the case of even $L=2n$ with distinct connectivities, it was 
conjectured in \PRdGN\ that 
the sum of components of the Perron--Frobenius eigenvector $\Phi^*$ 
over all loop
configurations $\pi$ is
\eqn\sumloopev{
\sum_\pi \Phi^*_\pi =A_{{\rm HT}}(2n)=2,10,140,5544
\ldots\qquad L=2n=2,4,6,8,\ldots }
The largest component was also conjectured to be
\eqn\psimloop{\Phi^*_{max}=A_{{\rm HT}}(2n-1)=1,3,25,588, 
\ldots\qquad L=2n=2,4,6,8,\ldots}

\fig{A sample of half turn symmetric FPL configurations of size 4 and 5.
The lower half of these configurations has a link pattern 
described by, respectively, $\{\frown\ \ \frown\}$, $\{\lharch\ \ \frown\ \ \rharch\}$
and $\{\frown\ \ \frown\ \ \cdot\}$.
}{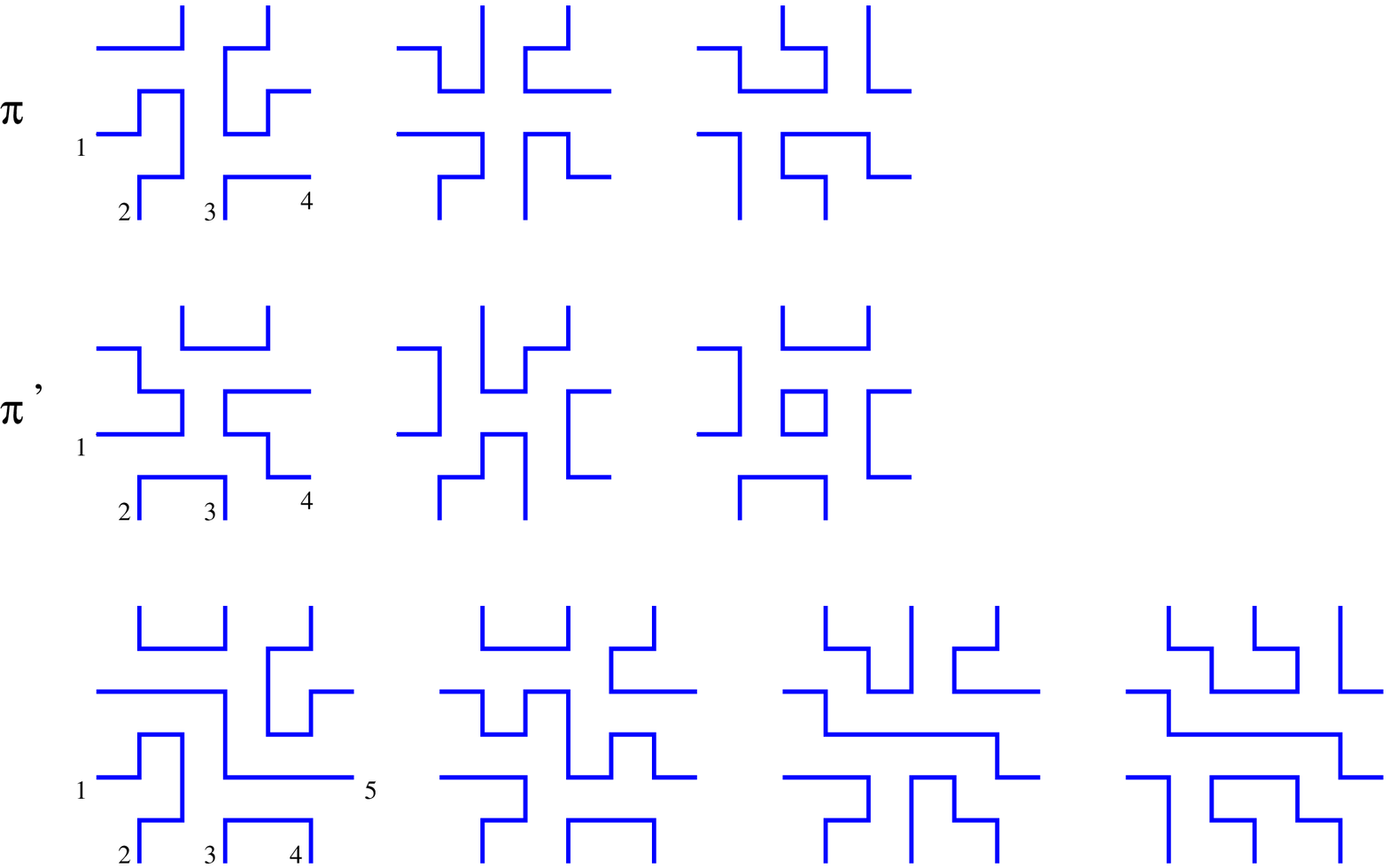}{8.cm}
\figlabel\htfpl

In both the odd and the even cases, it was further conjectured 
in \RStr\ and in \PRdGN, respectively,  that the individual components 
$\Psi_\pi$, resp. $\Phi^*_\pi$, 
count the number of HTSFPLs, that is, half-turn symmetric
FPL configurations drawn on a $L\times L$ grid, 
 whose connectivity pattern is described by the 
arch pattern $\pi$. For example, \par \noindent
the unique HTSFPL pertaining to $L=3$ and the link 
\par \vskip-9mm \noindent  
pattern $\pi=\{\frown\ \ \cdot\}$ is as shown here on the right: 
%\qquad
\hfill \hbox{\epsfxsize=15mm\epsfbox{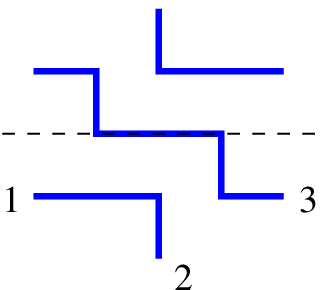}}\qquad,\qquad  \par\noindent
while  further examples for $L=4,5$ are depicted in Fig. \htfpl.

For completeness and future use in this paper
we also recall here the sum rule written in \BdGNun\ for the 
loop model of even size $L=2n$ with ``identified connectivities''.
The Perron--Frobenius eigenvector denoted $\Phi$ satisfies
\eqn\wellknownb{\sum \Phi_\pi=A_n\ ,}
together with $\Phi_{max}=A_{n-1}$. While the latter remains 
a conjecture,   \wellknownb\  has now been established \DFZJ\
as we recall below.

Turning now to the XXZ spin chain of odd size, 
the parallel (and prior!) conjectures read \RSun % [RS0012141]
\eqn\sumspin{\sum_\alpha \tilde\Psi_\alpha=3^{n/2} 
\CN_n=1,3,15,126,1782\ldots\qquad L=2n+1=
1,3\ldots}
for the sum over all spin configurations of total spin 1/2, the largest 
component is
\eqn\psimspin{\tilde\Psi_{max}=A_n=1,1,2,7,42\ldots\qquad L=2n+1=1,3,\ldots}
and the square norm of $\Psi$ is
\eqn\normsq{\sum_\alpha \tilde\Psi_\alpha^2=\CN_n^2=A_{{\rm HT}}(2n+1)\ }
when the normalization is such that $\tilde \Psi_{%\underbrace
{+\cdots+}%\underbrace 
{-\cdots-}}=1$.
For an even size $L=2n$, 
(with twisted boundary conditions, see 
Sect.~4),  the ground state wave function is complex, and 
 the parallel conjectures of \BdGNun\ and \RSunb\ read
\eqna\sumspinev
$$
\eqalignno{ \sum_\alpha\tilde\Phi_\alpha&= 3^{n/2} A_n &\sumspinev a\cr
\sum_\alpha \tilde\Phi_\alpha^2 &= A_n^2 &\sumspinev b\cr
\sum_\alpha |\tilde\Phi_\alpha|^2 &= A_{{\rm HT}}(2n) &\sumspinev c\cr
\tilde\Phi_{max}=\tilde\Phi_{(+-)^n}&=  \CN_{n-1} % <<<
\e{i\pi/6}&\sumspinev d\cr
}$$
with the normalization that $\tilde\Phi_{min}=\tilde\Phi_{++\cdots+--\cdots-}
=\e{i\pi n/6}$. The first of these conjectures is now established, since it is
a corollary of the sum rule \wellknownb, as will be reexplained
below.

{\it Examples:} $L=5$ ($n=2$): 
$\Psi=(1,1,1,1,1,4,4,4,4,4)$ in the loop basis, 
$\tilde\Psi=(1,1,1,1,1,2,2,2,2,2)$ after change
to the spin basis. 
Note that for $L$ odd, there are always $L$ repeats due to the breaking of
rotational symmetry by the defect. Indicating these repeats with a
superscript, at $L=7$,
$\Psi=(1^{(7)},6^{(7)},14^{(14)},49^{(7)})$ for link patterns, 
$\tilde\Psi=(1^{(7)},3^{(14)},4^{(7)},7^{(7)})$ for spins.
In even sizes, for $L=4$, $\Phi^*=(3^{(2)},1^{(4)})$ and
$\tilde\Phi=(\sqrt{3}e^{\pm i\pi/6},e^{\pm i\pi/3},1^{(2)})$; 
for $L=6$,
$\Phi^*=(25^{(2)},9^{(6)},5^{(6)},1^{(6)})$,
$\tilde\Phi=(5 e^{\pm i\pi/6},(\sqs e^{\pm i \alpha})^{(3)},
(\sqs e^{\pm i \beta})^{(2)}, \sqs e^{\pm i \gamma},
(e^{\pm i\pi/6})^{(2)},e^{\pm i\pi/2}
)$ with $\alpha=\Arctan{1/3\sqt}$,  $\beta={\pi\over 3}-\alpha$,
$\gamma={\pi\over 3}+\alpha$. 
%etc. % $\tilde\Phi=\ldots$. 

In view of previous experience, 
it is very natural to extend the discussion  to the 
inhomogeneous version of the loop model (whose precise definition  is
recalled below in Sect.~2). Indeed the main result of \DFZJ\ (Theorem 5) 
is that for the even size (IC) loop model, 
\eqn\wellknown{\sum_{\pi\in LP_{2n}}\Phi_\pi(z_1,\ldots,z_{2n})=3^{-n(n-1)/2} 
s_{\ppY}(z_1,\ldots,z_{2n})\ ,}
so that in the homogeneous limit $z_i\to 1$, \wellknownb\ follows. 
The main result of {\it this} paper is that 
\eqn\sumloopz{
\cases{\sum_{\pi\in LP_{2n+1}} \Psi_\pi(z_1,\ldots,z_{2n+1})= 
3^{-n^2} s_{\Y}(z) s_{\pY}(z)& if $L=2n+1$ \cr
\cr
\sum_{\pi\in LP^*_{2n}} \Phi^*_\pi(z_1,\ldots,z_{2n})= 
3^{-n(n-1)}\, s_{\pY}(z) s_{\ppY}(z) & if $L=2n$ \ ,\cr}}
as we will show in Sect.~2.5 and 3. 
In these expressions, $s_Y(z)$ stands for the Schur function 
labelled by a Young diagram $Y$, 
a symmetric function of $z_1,\ldots,z_{L}$. The diagrams 
 $\pY$ and $\ppY$ have been defined above, after \formOk, thus 
$\Y$ has two rows of length $n$, two of length $n-1$, etc, two of length
1. From this follow the proofs of the sum rules
\sumloop\ and \sumloopev. 
The sum rule \sumspin\ for the odd spin chains 
will be also derived in Sect.~4.5, and \normsq\ and \sumspinev{b} in Sect.~4.6.
On the conceptual level, it may be  interesting to notice that 
our derivation makes use not only of the previously mentionned
techniques, introduction of spectral parameters,  recursion equations and
$q$KZ equation, etc, but also of a new idea borrowed from knot theory, 
namely the use of skein relations. 

The educated reader will recognize in \sumloopz\ formulae equivalent 
to those written by Razumov and Stroganov \RSqu\ for the partition 
function of the square-ice model with boundary conditions appropriate
to the enumeration of half-turn symmetric ASMs. We return to this 
in our  Conclusion.

For the sake of the reader, we summarize in Table 1 and Table 2
some notations and
data for the various situations that we consider in this paper.

\bigskip
\centerline{\vbox{
\halign{ \hfil #\hfil\quad  && \hfil # \hfil\cr
 Size $L$ & Model & Space & dimension &\vtop{\hbox{Perron--Frobenius}\hbox{eigenvector}}&
\vtop{\hbox{Transfer}\hbox{matrix}}\cr
\noalign{\vskip5pt}
$2n+1$ & loop & $LP_{2n+1}$ & $2n+1\choose n$ & $\Psi_\pi$&$T$\cr
$2n$ & loop (``I.C.'') & $LP_{2n}$ & $C_n={(2n)!\over n!(n+1)!}$ & 
                                                        $\Phi_\pi$&$T$\cr
$2n$ & loop (``D.C.'') & $LP^*_{2n}$ & ${2n\choose n}$ & $\Phi^*_\pi$&$T^*$\cr
$2n+1$ & spin & $ \{s_z=\oh\} $ & $C_{n+1}$ & $\tilde\Psi_\alpha$&$\tilde T$\cr
$2n$ & spin & $\{s_z=0\} $ & $C_n$ & $\tilde\Phi_\alpha$&$\tilde T$\cr
}}}
\smallskip
\centerline{{\bf Table 1}. Notations for the different models and boundary conditions considered in this paper}
\penalty -5000

\hskip1cm\vbox{
\halign{ \hfil #\hfil\quad  & \hfil # \hfil
& \hfil  # \hfil & \hfil # \hfil& \hfil# \hfil
\cr   
$B$ & equ. (4.10)  
& \qquad\qquad &
$B_0,\ B_\infty$ & equ. (3.2) \cr
$L_{2n+1}$ & \S~2.1
& \qquad\qquad &
 $L_{2n+2}$ &\S~ 2.3 \cr
 $ P_0,\ P_\infty$ & \S~2.3 
& \qquad\qquad &
$Q,\ P_\pm $ & \S~4.3 \cr
$S$ & \S~4.2 
& \qquad\qquad &
$s_{Y_{n}}\,,\ s_{Y'_{n}}$, $Y_{n}\,,\ Y'_{n}$
& equ. (1.3), (1.14)\cr
}}
\smallskip
\centerline{{\bf Table 2}. Some other notations with the section or equation
number where they first appear.}

%%%%%%%%%%%%%%%%%%%%%%%%%%%%%%%%%%%%%%%%%%%%%%%%%%%%%%%%%%%%%%%%%%%%%%%%%%%%
\def\P{{\rm P}}

\newsec{The inhomogeneous $O(1)$ loop model in odd size}

\subsec{Link patterns, transfer matrix and $R$-matrices}
\noindent
The model is defined on a semi-infinite cylinder of square 
lattice of odd perimeter $2n+1$
whose faces are covered by either of the two following face configurations
\eqn\faceconf{\vcenter{\hbox{\epsfbox{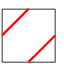}}} \quad {\rm or} \quad \vcenter{\hbox{\epsfbox{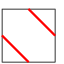}}}\ .}
In a given configuration, labelling cyclically $i=1,2,\ldots 2n+1$ 
the centers of the boundary edges (and with 
the convention that $i+2n+1\equiv i$),
we note that these points are connected among themselves or to the 
point at infinity via nonintersecting curves.
Actually, drawing on each face the two configurations \faceconf\ with probabilities $p$ and $1-p$,
$p\in (0,1)$, leads only to situations where {\it one}\/ boundary point is connected to infinity,
while the remaining $2n$ are pairwise connected. 
Forgetting about the underlying lattice, the
connection pattern, also called link pattern, is simply a chord diagram, namely a configuration of 
$2n+1$ points on a circle, pairwise connected by 
nonintersecting arcs within the interior disc,
with one unmatched point that we connect to the center of the disc.
The set of such link patterns is denoted 
by $LP_{2n+1}$ and has cardinality 
$d_n={2n+1 \choose n}$.  Later on, we regard this set as a linear space,
 spanned by the previous link patterns, thus of dimension $d_n$. 

Here we study the inhomogeneous version of this model, in which plaquettes above the boundary point
$i$ are picked among the two faces \faceconf\
 with respective probabilities $p_i$ and $1-p_i$, $p_i\in (0,1)$. We wish to compute the probabilities 
${\P}_\pi$ that random configurations connect the boundary point according 
to a given $\pi\in LP_{2n+1}$.
Clearly, these probabilities are invariant under the addition of a row of $2n+1$ plaquettes to
the original semi-infinite cylinder, which amounts to an equation of the form
\eqn\transma{ \sum_{\pi'\in LP_{2n+1}}T(p_1,\ldots, p_{2n+1})_{\pi,\pi'} {\P}_{\pi'}={\P}_\pi }
where the transfer matrix $T$ acts on link patterns in an obvious way, by concatenation. The vector
${\P}=\{ {\P}_\pi\}_{\pi\in LP_{2n+1}}$ is determined as the properly 
normalized Perron--Frobenius eigenvector of $T$, 
with eigenvalue $1$.

This sytem is known to be integrable, as $T$ may be constructed by multiplying and then tracing
plaquette operators that satisfy the Yang--Baxter equation. Indeed, parametrizing the probabilities as
\eqn\paraprob{p_i={q\, z_i - q^{-1} t\over q\, t -q^{-1} z_i} }
where $q$ is a complex cubic root of unity $q=-e^{i\pi/3}$, we may identify the $i$-th plaquette operator
(or $R$-matrix) as 
\eqn\rmat{ R_i(z_i,t)={q\, z_i - q^{-1} t\over q\, t -q^{-1} z_i}\ \vcenter{\hbox{\epsfbox{face2.eps}}} \ 
+ {z_i -t\over q\, t -q^{-1} z_i}\ \vcenter{\hbox{\epsfbox{face1.eps}}} }
where $z_i$ and $t$ are spectral parameters attached respectively to the vertical line above point $i$,
and to the horizontal one running around the cylinder. With the parametrization \paraprob, the probability
vector $\P$ is clearly a rational fraction of the $z$'s. In the sequel, we will use a different
normalization $\Psi \propto {\P}$ in which the entries of $\Psi$ are coprime polynomials of the $z$'s.

In \refs{\DFZJ,\DFZJb}, it was shown that the transfer matrix relation \transma\ may be equivalently replaced by 
a system of relations of the form
\eqn\syseq{ \Rc_{i,i+1}(z_i,z_{i+1}) {\Psi} = \tau_i {\Psi} , \ i=1,2,\ldots 2n+1}
where $\tau_i$ is simply the interchange of spectral parameters $z_i\leftrightarrow z_{i+1}$
and $\Rc$ is the tilted plaquette operator
\eqn\defcr{ \Rc_{i,i+1}(z,w)= {q\, z - q^{-1} w\over q\, w -q^{-1} z} I + {z -w\over q\, w -q^{-1} z} e_i}
in terms of the local identity $I$ and Temperley--Lieb operators $e_i$, $i=1,2,\ldots ,2n+1$
defined pictorially as
$$I=\vcenter{\hbox{\epsfbox{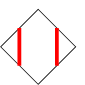}}}\qquad {\rm and}
\qquad e_i=\vcenter{\hbox{\epsfbox{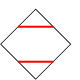}}}$$
and acting at points $i$ and $i+1$ on link patterns by concatenation as indicated schematically below:
\eqn\actei{ \vcenter{\hbox{\epsfbox{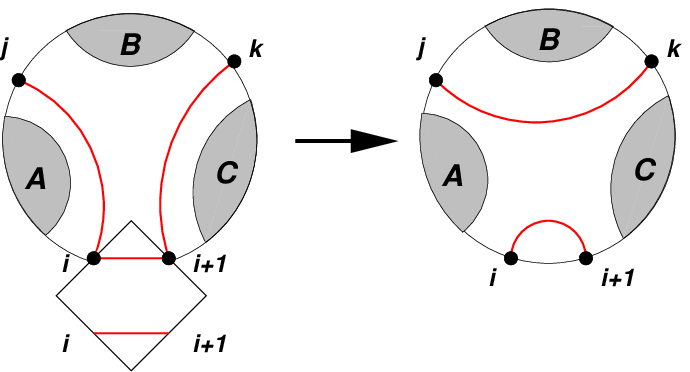}}} }
Note that if $k=i$ and $j=i+1$, $e_i$ acts by creating a loop which we allow ourselves to erase, and therefore
this action leaves the pattern invariant.

The $e_i$ obey the usual conditions
$e_ie_{i\pm 1}e_i=e_i$ and $e_i^2=e_i$ that define the (cyclic) Temperley--Lieb algebra $TL(1)$.

The last equation $i=2n+1$
of \syseq\ may be replaced by a cyclic invariance condition
\eqn\cycinv{  {\Psi}(z_2,z_3,\ldots,z_{2n+1},z_1)={\Psi}(z_1,z_2,
\ldots,z_{2n+1})\ . }

Finally, we will also be using $\Rc$ matrices with the second 
spectral parameter sent
either to zero or to infinity. Up to a multiplicative redefinition by $-q^{\pm 3/2}$,
the two corresponding plaquette operators\foot{Note that the following choice of normalization 
for $t_i$ is ad-hoc
to ensure rotational invariance of the crossing move, and coincides with the standard crossing
operators of knot theory. However, the prefactors $-q^{\pm 3/2}$ will be irrelevant at the
particular point $q^{1/2}=e^{-i\pi/3}$ to which we will restrict later on.}
\eqn\skeinrela{\eqalign{ t_i^{-1}&\equiv -q^{3/2}\Rc_{i,i+1}(z,0)= q^{-1/2} I +q^{1/2} e_i \cr
t_i&=-q^{-3/2}\Rc_{i,i+1}(z,\infty)=q^{1/2} I +q^{-1/2} e_i \cr}}
may be interpreted respectively as under- and over-crossings of links, with pictorial representation
$$t_i=\vcenter{\hbox{\epsfbox{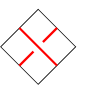}}}\qquad {\rm and}
\qquad t_i^{-1}=\vcenter{\hbox{\epsfbox{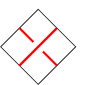}}}$$
and the two pieces of Eq.~\skeinrela\ are nothing else than the celebrated skein relations for knots, 
which read pictorially
\eqn\pictoskein{\eqalign{ \vcenter{\hbox{\epsfbox{undercross.eps}}} &= 
q^{-1/2}\ \vcenter{\hbox{\epsfbox{id.eps}}} \ \ +q^{1/2} \ \vcenter{\hbox{\epsfbox{ei.eps}}} \cr
\vcenter{\hbox{\epsfbox{overcross.eps}}}&= 
q^{1/2}\ \vcenter{\hbox{\epsfbox{id.eps}}} \ \ +q^{-1/2} \ \vcenter{\hbox{\epsfbox{ei.eps}}}\ .\cr}}

%%%%%%%%%%%%%%%%%%%%%%%%%%%%%%%%

\subsec{The quantum Knikhnik--Zamolodchikov equation}
\noindent
More generally, it was noted in \Pas\ that the condition that 
$q$ be a cubic root of unity could be relaxed
and that equations \syseq\ and \cycinv\ are a particular case 
of quantum Knizhnik--Zamolodchikov ($q$KZ)
 equation \refs{\DFZJc,\DFb}. The latter amounts to the system
\eqn\qkz{\eqalign{
\Rc_{i,i+1}(z_i,z_{i+1}) \Psi &= \tau_i \Psi\qquad \ i=1,2,\ldots 2n\cr
\sigma \Psi(z_2,\ldots,z_{2n+1},s z_1)&=c \Psi( z_1,\ldots,z_{2n+1})\cr}}
where $s,c$ are scalars and the operator $\sigma$ acts on link patterns via the cyclic rotation
of labels $i\to i+1$. Here $\Rc$ denotes the same plaquette operator as before, except 
that its definition involves the generators $e_i$ of the Temperley--Lieb algebra $TL(\tau)$, 
% >>
({\it i.e.} satisfying the relation $e_i^2=\tau e_i$),
with $\tau=-q-q^{-1}$.
Note that we may replace the second line of \qkz\ by an equation of the form
\eqn\cycr{ \Rc_{2n+1,1}(z_{2n+1},s z_1)\Psi(z_1,\ldots,z_{2n+1}) =\Psi(s^{-1}z_{2n+1},\ldots,sz_1)}
where the omitted $z$'s are left unchanged.

We now look for polynomial solutions $\Psi(z_1,\ldots,z_{2n+1})$ of this system, with {\it minimal}
degree. From the first line of Eq.~\qkz\ we learn that whenever points $i$ and $i+1$ are not 
connected in $\pi$, $\Psi_\pi$ factors out a term $qz_i-q^{-1} z_{i+1}$, and more generally
if no two points between $i$ and $j>i$ are connected, then $\Psi_\pi$ factors out $qz_i-q^{-1}z_j$.
% >>
For the link pattern $\pi_0$ connecting points $i \leftrightarrow 
2n+2-i$, $i=1,2,\ldots,n$ while $n+1$ is unmatched, this fixes the 
base component $\Psi_{\pi_0}$ to be
\eqn\tobe{ \Psi_{\pi_0}=\prod_{1\leq i<j \leq 2n+1} {q\,z_i-q^{-1}z_j\over q-q^{-1}} }
up to multiplication by a symmetric polynomial, which we pick to be $1$ for the sake of minimality.
This in turn determines both scalars in \qkz\ to be $s=q^3$ and $c=q^{3n}$.
The other entries of $\Psi$ are then determined from Eq.~\qkz, which reads in components
\eqn\compo{\eqalign{ -(q^{-1}z_{i+1}-qz_i)\partial_i \Psi_\pi&=
\sum_{\pi'\neq \pi\atop e_i\pi'=\pi} \Psi_{\pi'}\cr
\Psi_\pi(z_2,\ldots,z_{2n+1},s z_1)&=c \Psi_{\sigma^{-1}\pi}
(z_1,\ldots,z_{2n+1})\ ,\cr}}
where 
\eqn\defpartial{
\partial_i f(z_1,\ldots,z_i,z_{i+1},\ldots,z_{2n+1})
:={f(z_1,\ldots,z_{i+1},z_i,\ldots,z_{2n+1})-f(z_1,\ldots,z_i,z_{i+1},\ldots,z_{2n+1})\over z_{i+1}-z_i}\ .} % <<<

Also true for any $q$ is the recursion relation:
\eqn\recupsi{\eqalign{\Psi_{\phi_i(\pi)}(z_1,\ldots,z_L)&|_{z_{i+1}=q^2 z_i}
={1\over (q-q^{-1})^{2L-3}}z_i\prod_{j=1}^{i-1} (q z_j-q^{-1}z_i) (z_j-q z_i)\cr
&\times 
\prod_{j=i+2}^L (q z_{i+1}-q^{-1}z_j) (z_{i+1}-q z_j)
\Psi_\pi(z_1\ldots z_{i-1},z_{i+1}\ldots z_L)\cr}}
where $\phi_i$ inserts two consecutive points connected via a ``little arch" between
points $i-1$ and $i$ in any link pattern $\pi$ of size $2n-1$; all other components vanish
when $z_{i+1}=q^2 z_i$.

\subsec{Projections to link patterns of larger size}
\noindent
In this section, we introduce two projection operators $P_0$ and $P_\infty$ 
from $LP_{2n+1}$ into 
the vector space generated by the link patterns with $2n+2$ points without unmatched points,
denoted $LP_{2n+2}$. These two projections would allow us
to relate the solutions of the $q$KZ equation for 
punctured discs with $2n+1$ points on their perimeter 
to those for unpunctured discs of perimeter $2n+2$. 
We decide from now on to restrict ourselves to the particular value $q=-e^{i\pi/3}$
(RS point), as we will be mostly reasoning on transfer matrices
 for the loop model on a punctured disk, for which the periodicity is 
requested.

\fig{Example of projections $P_0$ and $P_\infty$ 
of a link pattern of size $13$.
We first add a point $14$ between $1$ and $13$ on the boundary of the link pattern, and then
connect the latter to the center, itself linked to the 
unmatched point (labelled $5$ here), via a link passing
under ($P_0$) or over ($P_\infty$) those separating the two points. Finally, we must use the
skein relations \skeinrela\ to express the latter as linear combinations of non-crossing
link patterns of size $14$.}{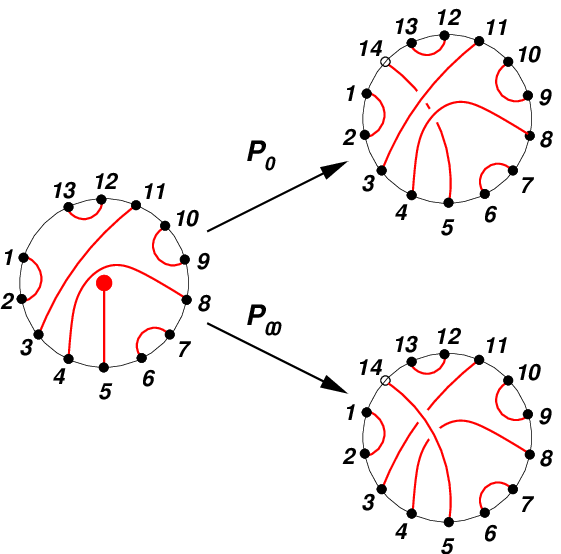}{5.5cm}
\figlabel\proj

The projections $P_0$ and $P_\infty$ are defined as follows.
For any link pattern of size $2n+1$, let us first add an extra point labelled $2n+2$ on
the boundary of the disc, between points $1$ and $2n+1$. Next 
this point is linked to the center of the disc, which is itself connected
to the (unique) unmatched point of the link pattern. But by doing so, 
the added link may have to cross
existing links. As illustrated in Fig.\proj, we define $P_0$ by imposing that all these 
crossings be undercrossings, and $P_\infty$ by imposing that they all be overcrossings.
We then simply have to use the skein relations \pictoskein\ at each crossing to obtain
a linear combination of non-crossing link patterns of size $2n+2$. 

It is also clear that the mappings $P_0$ and $P_\infty$ are
surjective: every $\pi'$  in $LP_{2n+2}$ in which the point $2n+2$ is
matched with some $i$ is the image by $P_0$ or $P_\infty$
of a $\pi\in LP_{2n+1}$ in which $i$ is unmatched and the other points
form the same pairs. This justifies to call $P_0$ and $P_\infty$ 
``projections''  from  $LP_{2n+1}$ to $LP_{2n+2}$.

For illustration, there are $3$ link patterns of size $3$, and their projections in the vector space
of non-crossing link patterns of size $4$ read:
\eqn\projexo{ \eqalign{
P_0\ \ \vcenter{\hbox{\epsfbox{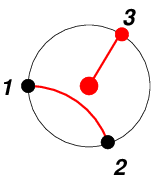}}} &=\ \vcenter{\hbox{\epsfbox{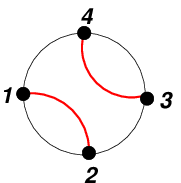}}} \cr
P_0\ \ \vcenter{\hbox{\epsfbox{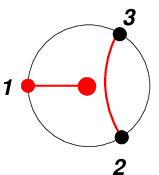}}} &=\ \vcenter{\hbox{\epsfbox{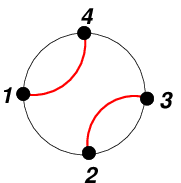}}}\cr
P_0\ \ \vcenter{\hbox{\epsfbox{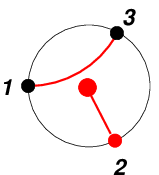}}} &= q^{-1/2} \ \vcenter{\hbox{\epsfbox{lipa4-1.eps}}}
+q^{1/2} \ \vcenter{\hbox{\epsfbox{lipa4-2.eps}}}\cr}}

\eqn\projexinf{ \eqalign{
P_\infty\ \ \vcenter{\hbox{\epsfbox{lipa3-1.eps}}} &=\vcenter{\hbox{\epsfbox{lipa4-1.eps}}} \cr
P_\infty\ \ \vcenter{\hbox{\epsfbox{lipa3-2.eps}}} &=\vcenter{\hbox{\epsfbox{lipa4-2.eps}}} \cr
P_\infty\ \ \vcenter{\hbox{\epsfbox{lipa3-3.eps}}} &=  q^{1/2} \ \vcenter{\hbox{\epsfbox{lipa4-1.eps}}}
+q^{-1/2} \ \vcenter{\hbox{\epsfbox{lipa4-2.eps}}}\cr}}

\subsec{Projections and Transfer Matrices}
\noindent
Let $T(t|z_1,\ldots,z_{2n+1})$ denote the transfer matrix for the loop model on a 
{\it punctured disk} of perimeter $2n+1$, obtained from \transma\ via the parametrization 
\paraprob, and likewise $T(t|z_1,\ldots,z_{2n+2})$ for a {\it non-punctured}
disk of perimeter $2n+2$. Then the projection operators $P_0$ and $P_\infty$ of previous 
section act as intertwiners between odd and even transfer matrices, where in the 
latter the additional spectral parameter is taken to $0$ or $\infty$ respectively, namely:
\eqnn\projtm
$$\eqalignno{
P_0 T(t|z_1,\ldots,z_{2n+1})&=T(t|z_1,\ldots,z_{2n+1},0) P_0\cr
P_\infty T(t|z_1,\ldots,z_{2n+1})&=T(t|z_1,\ldots,z_{2n+1},\infty) P_\infty
\ .&\projtm\cr
}$$

\fig{The projection $P_0$ intertwines the transfer
matrices for size $2n+1$ and $2n+2$, the latter with $z_{2n+2}=0$.
We have represented the transfer matrices pictorially as adding an extra row to the
disk. Each intersection between lines stands for an $\Rc$ matrix
operator. That $z_{2n+2}=0$ imposes here that the added
intersection be an undercrossing, in agreement with the action of $P_0$.}{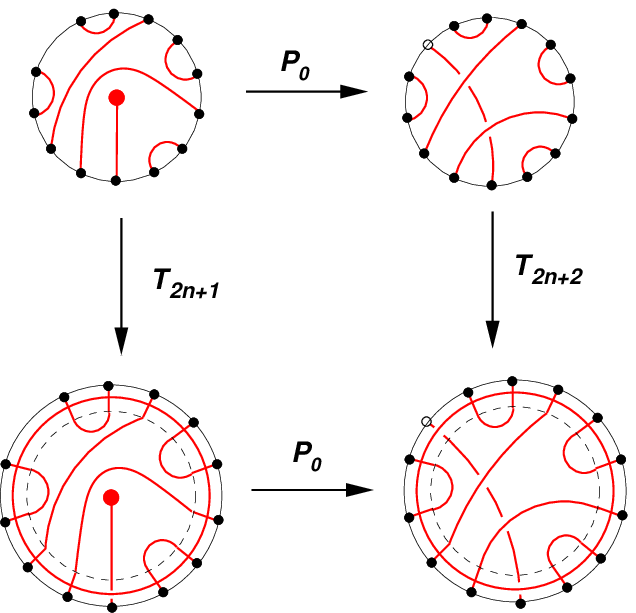}{8cm}\figlabel\projtmfig
This equality is easily proved graphically, see Fig.~\projtmfig.

Apply now Eqs.~\projtm\ to $\Psi$: we find that $P_\bullet \Psi$ is an eigenvector 
of $T(t|z_1,\ldots,z_{2n+1},\bullet)$
with eigenvalue $1$, $\bullet=0,\infty$. This eigenvalue being generically non-degenerate, we conclude
that $P_\bullet \Psi$ is proportional to the vector $\Phi(z_1,\ldots,z_{2n+1},\bullet)$, 
defined as the ground state eigenvector of the system on a disk of even perimeter $2n+2$, 
as introduced in \DFZJ:
\eqnn\polmulb
$$
\eqalignno{P_0\Psi&=A_0(z_1,\ldots,z_{2n+1})\, \Phi(z_1,\ldots,z_{2n+1},0)\cr
P_\infty\Psi&=A_\infty(z_1,\ldots,z_{2n+1})\, 
\Phi(z_1,\ldots,z_{2n+1},\infty)\ . &\polmulb\cr}
$$
Here the notation $\Phi(z_1,\ldots,z_{2n+1},\infty)$ is slightly abusive, and stands
for the limit $\lim_{z_{2n+2}\to \infty} \Phi(z_1,\ldots,z_{2n+1},z_{2n+2})/z_{2n+2}^n$.
The polynomial character of the quantities $A_0$ and $A_\infty$ may be deduced from the recursion
relations \recupsi\ by sending spectral parameters to $0$ or $\infty$; these polynomials
will be determined in the next section.

\subsec{Ground state sum rule}
\noindent
The fundamental remark of Razumov and Stroganov is 
that at the special point $q=-e^{i\pi/3}$ (RS point)
the problem simplifies drastically, and leads to all 
sorts of combinatorial wonders. This point 
may be characterized as the unique one (up to conjugation) where there exists a non-trivial
{\it common} left eigenvector to all operators $e_i$, or equivalently to all plaquette operators $\Rc$.
In view of the action \actei\ of $e_i$ on link patterns, it is clear that, under $e_i$, each link pattern
gives rise to exactly one link pattern. However, if $i$ is already connected to $i+1$ in the link pattern,
the action of $e_i$ creates a loop, which may be removed at the expense of a multiplicative factor
$\tau$. The existence of a common left eigenvector to all $e_i$ imposes therefore that $\tau=1$,
for which we pick the root  $q=-e^{i\pi/3}$, $q^{1/2}=e^{-i\pi/3}$. 
This eigenvector is then simply the sum of components over all link patterns, namely the 
covector $v$ with all entries equal to $1$ in the link pattern basis.

As a consequence of this definition, we have $v\Rc=v$ for all arguments, 
including when the second spectral
parameter is zero or the infinity, hence $vt_i=vt_i^{-1}=v$ as 
well (as $q^{3/2}=-1$ at the RS point). 
Moreover, let $v'$ be the vector with all entries $1$ in the link pattern basis of
$LP_{2n+2}$. This implies that
\eqn\invsums{v' P_0 = v\quad {\rm and} \quad v'P_\infty = v}
hence we finally get
\eqn\sumrul{\eqalign{ v\cdot \Psi&=A_0(z_1,\ldots,z_{2n+1}) v'\cdot \Phi(z_1,\ldots,z_{2n+1},0)\cr
v\cdot \Psi&=A_\infty(z_1,\ldots,z_{2n+1}) v'\cdot \Phi(z_1,\ldots,z_{2n+1},\infty)\ .\cr}}
In \DFZJ, 
the quantity $v'\cdot \Phi(z_1,\ldots,z_{2n+2})$ has been identified with the so-called 
Izergin--Korepin determinant,
which, at the RS point, is equal to the GL$(2n+2)$ Schur function 
$s_\Y(z_1,\ldots,z_{2n+2})$ for the Young diagram 
$\Y$ with two rows of $n$ boxes, two rows of $n-1$ boxes, 
etc..., two rows of one box, as defined in Sect.~1. 
Sending $z_{2n+2}$ to $0$ amounts to restricting the Schur function to 
GL$(2n+1)$, while keeping the
same tableau $\Y$. Sending $z_{2n+2}$ to $\infty$ amounts to 
restricting the Schur function to GL$(2n+1)$,
while truncating the Young tableau into $\pY$ equal to $\Y$ with 
the first row removed.
Hence $v\cdot \Psi=A_0 s_\Y(z_1,\ldots,z_{2n+1})=A_\infty s_\pY(z_1,\ldots,z_{2n+1})$.
Finally, we make the hypothesis that the Schur functions 
$s_\Y(z_1,\ldots,z_{2n+1})$ and $s_\pY(z_1,\ldots,z_{2n+1})$
have no common factor. From this hypothesis, 
which is supported by the examination of the first cases
but would require a complete proof, we deduce that $v\cdot \Psi$ 
is a polynomial multiple of $s_\Y s_\pY$. Now the degree of 
$s_\Y$ is the total number of boxes in
$\Y$, hence $s_\Y s_\pY$ has degree $n(n+1)+n^2=n(2n+1)$. 
This coincides with the degree of $\Psi$ 
as given by that of $\Psi_{\pi_0}$  in Eq..~\tobe, hence we conclude that $v\cdot\Psi$ is proportional
to $s_\Y s_\pY$ by a scalar factor, fixed to be $3^{-n^2}$ by the recursion relation \recupsi. 

We therefore identify the polynomials $A_0\propto s_{\pY}$ and $A_\infty\propto
s_\Y$, and get the multi-parameter sum rule
\eqn\multisum{ \sum_{\pi \in LP_{2n+1}} \Psi_\pi(z_1,\ldots,z_{2n+1})=3^{-n^2} s_\Y(z_1,\ldots,z_{2n+1}) 
s_{\pY}(z_1,\ldots,z_{2n+1})\ . }
In the homogeneous limit where all the $z_i$ tend to 1, the Schur functions 
$s_\Y$ and $s_{\pY}$
reduce to the dimensions of the corresponding representations of GL($2n+1$), and we note that 
$\dim_\Y=\dim_{\pY}$ because the two Young diagrams form together a rectangle of size
$(2n+1)\times n$, %. 
(and thus $s_\pY({1\over z})=\prod_j z_j^{3n}s_\Y(z)$).
 This leads to \sumloop\ via \formOk. 

%%%%%%%%%%%%%%%%%%%%%%%%%%%%%%%%%%%%%%%%%%%%%%%%%%%%%%%%%%%%%
\newsec{The inhomogeneous $O(1)$ loop model in even size}
\noindent
As discussed in Sect.~1, the inhomogeneous $O(1)$ loop model may also be considered
on a punctured disk of even perimeter $2n$:  
link patterns now have an isolated puncture in the center of the disk, and we denote
by $LP_{2n}^*$ their set, of cardinality $2n\choose n$. 
Applying the same line of thought as in the previous section, 
we now define mappings $P_0$ and $P_\infty$ from   $LP_{2n}^*$
to  $LP_{2n+1}$, by simply creating an extra point between the points $1$ and $2n$, 
and connecting it to the 
puncture via a link passing {\it below} or {\it above} all the crossed ones. 
Comparing the dimensions it is clear that $P_0$ and $P_\infty$ cannot be surjective, though they
are presumably injective. Using the same type of arguments as in Sect.~2, we find that there
is an intertwining relation
\eqn\eveninter{\eqalign{
P_0 T^*(t|z_1,\ldots,z_{2n})&=T(t|z_1,\ldots,z_{2n},0) P_0\cr
P_\infty T^*(t|z_1,\ldots,z_{2n})&=T(t|z_1,\ldots,z_{2n},\infty) P_\infty
\ . \cr}}
where $T^*$ is the inhomogeneous transfer matrix acting on the span of $LP_{2n}^*$.
This now leads  to
\eqn\evenrel{\eqalign{
B_0(z_1,\ldots,z_{2n}) P_0\Phi^*(z_1,\ldots,z_{2n})&= \Psi(z_1,\ldots,z_{2n},0)\cr
B_\infty(z_1,\ldots,z_{2n})P_\infty\Phi^*(z_1,\ldots,z_{2n})&= \Psi(z_1,\ldots,z_{2n},\infty)\cr}}
where once again $\Psi(z_1,\ldots,z_{2n},\infty)$ denotes the highest degree
($=2n$)  terms in $z_{2n+1}$ of $\Psi$ and $B_0$ and $B_\infty$ will be 
determined soon.

As a consequence, we find
that the even case sum rule $B_0 v\cdot \Psi(z_1,\ldots,z_{2n})$ 
must be equal to the quantities \multisum\ taken for $z_{2n+1}=0$ 
while $B_\infty v\cdot \Psi(z)$ equals that for $z_{2n+1}=\infty$ 
(with the abovementioned
division by an appropriate power of $z_{2n+1}$). These restrictions on the last spectral parameter
have the effect of truncating the Schur functions. More precisely, 
when $z_{2n+1}\to 0$, the factor $s_\Y(z_1,\ldots,z_{2n+1})$ tends to
$s_{\Y}(z_1,\ldots,z_{2n})=(z_1\ldots z_{2n}) s_{\ppY}(z_1,\ldots,z_{2n})$ 
where $\ppY$ now has
two rows of length $n-1$, two rows of length $n-2$, etc. two rows of length $1$. Meanwhile, the factor 
$s_{\pY}(z_1,\ldots,z_{2n+1})$ tends to $s_{\pY}(z_1,\ldots,z_{2n})$.
When $z_{2n+1}\to \infty$, the leading coefficient in $s_\Y(z_1,\ldots,z_{2n+1})$ truncates to
$s_{\pY}(z_1,\ldots,z_{2n+1})$ as explained before, while that in 
$s_{\pY}(z_1,\ldots,z_{2n+1})$
truncates to $s_{\ppY}(z_1,\ldots,z_{2n})$. This is consistent for
$B_0\propto z_1\ldots z_{2n}$ and $B_\infty$ a constant. 
In both limits, we reach the same sum rule
\eqn\evensum{ \sum_{\pi \in LP_{2n}^*} \Phi^*_\pi(z_1,\ldots,z_{2n}) 
=3^{-n(n-1)} s_{\ppY}(z_1,\ldots,z_{2n})\, s_{\pY}(z_1,\ldots,z_{2n})\ .}
The proportionality factor is fixed by the even analogue of the recursion relation
\recupsi, together with the normalization of the component 
\eqn\fundaeven{\Phi^*_{\pi_0}(z_1,\ldots,z_{2n})=\prod_{1\leq i<j\leq 2n}{q\,z_i-q^{-1}z_j\over
q-q^{-1}}} 
for the fully nested link pattern $\pi_0$ connecting points $i$ and $2n+1-i$, $i=1,2,\ldots,n$, 
while the puncture sits in the face delimited
by the little arch connecting $n$ to $n+1$ and the disk boundary.

%%%%%%%%%%%%%%%%%%%%%%%%%%%%%%%%%%%%%%%%%%%%%%%%%%%%%%%%%%%%%5
\newsec{The XXZ spin chain and six-vertex model at $\Delta=-1/2$}
\noindent We present here results concerning 
another closely related model: the XXZ spin chain at the value $\Delta={q+q^{-1}\over2}=-1/2$
of the anisotropy. As above we shall need an inhomogeneous version of
the model, which is the inhomogeneous six-vertex model.
New results are found mostly in the case of an odd size chain, since
the even case is already covered (if a little implicitly) by \DFZJ, 
but we shall need to introduce the model for arbitrary size anyway.

%%%%%%%%%%%%%%%%%%%%%%%%%%%%%%%%%%%%%%%%%%%%%%%%%%%%%%%%%%%%%%%%%%%%

\subsec{Definition of the XXZ spin chain}
\noindent The XXZ spin chain is given by the Hamiltonian
\eqn\HXXZ{
\tilde H=-{1\over 2}\sum_{i=1}^L (\sigma_i^x\sigma_{i+1}^x+\sigma_i^y\sigma_{i+1}^y
+\Delta \sigma_i^z\sigma_{i+1}^z)
}
acting on $({\Bbb C}^2)^{\otimes L}$, each ${\Bbb C}^2$ 
being a single spin space:
$\ket{+} \equiv\left( {1\atop 0}\right)$,
$\ket{-}\equiv \left({0\atop 1}\right)$. 
We first take $\Delta$ and $q$ generic, $\Delta={q+q^{-1}\over2}$. 
We choose here periodic boundary conditions for $L$ odd:
$\sigma_{L+1}\equiv \sigma_1$, 
and twisted periodic boundary conditions for $L$ even:
$\sigma^z_{L+1}\equiv \sigma^z_1$, $\sigma^\pm_{L+1}\equiv q^{\pm 2} \sigma^\pm_1$
where $\sigma^\pm=\sigma^x\pm i\sigma^y$. We can also write these as
$\sigma_{L+1}\equiv \Omega\, \sigma_1\, \Omega^{-1}$ with $\Omega=\left({-q\atop0}{0\atop -q^{-1}}\right)$ for $L$ even and $\Omega=1$ for $L$ odd.

We also define the transfer matrix of the inhomogeneous six-vertex model (see also
appendix B of \DFZJ\ where the case $L$ even is treated). For spectral parameters
$z_1,\ldots,z_L$, it is given by
\eqn\Tsixv{
\tilde{T}\equiv{\tilde T}(t|z_1,\ldots,z_L):=\tr_0 \left(R_{L,0}(z_L,t)\cdots R_{1,0}(z_1,t)\Omega\right)
}
where the $R$ matrices act on the tensor product of the physical space
$({\Bbb C}^2)^{\otimes L}$, each factor being labelled by $i=1,\ldots,L$, and of another ${\Bbb C}^2$,
labelled by $0$. 
The expression of $R$ acting in ${\Bbb C}^2\otimes {\Bbb C}^2$ reads (in the so-called 
homogeneous gradation)
\eqn\Rspin{
R(z,t)={1\over q\,t-q^{-1}z}\pmatrix{q\,z-q^{-1}t &0&0&0\cr 0&z-t&(q-q^{-1})t&0\cr 0&(q-q^{-1})z&z-t&0\cr 0&0&0&q\,z-q^{-1}t}\ .
}
The twist $\Omega$ acts on the auxiliary space,
%and is $1$ for $L$ odd, $\left({-q\atop0}{0\atop -q^{-1}}\right)$
%for $L$ even, 
which is consistent with the chosen boundary conditions for the spin chain since
$\Omega^{-1} R_{1,0}\Omega=R(\sigma_1,\Omega^{-1} \sigma_0\Omega)=R(\Omega\sigma_1\Omega^{-1},\sigma_0)
=R_{L+1,0}$ where the second equality follows from $U(1)$ invariance of the $R$-matrix.

When all $z$'s are equal, $\tilde{T}$ commutes with the Hamiltonian $\tilde H$. In particular
their ground states are identical.

%%%%%%%%%%%%%%%%

\subsec{Equivalence with the $O(1)$ loop model}
\noindent
For any $q$ there is a mapping from the $O(1)$ loop model to 
the XXZ/six-vertex model. 
% >> addition 
See also \Mitretal\ for a similar construction.
Call $\omega^{1/2}$ a square root of $\omega:=-q$.
To a link pattern $\pi$ in $LP_L$ associate the 
tensor product over
the set of arches of $\pi$, of the vectors $\omega^{1/2} \ket{+}_j\otimes \ket{-}_k
+\omega^{-1/2} \ket{-}_j\otimes \ket{+}_k$, where the indices $j$, $k$ are the endpoints of the arch (and
indicate the labels of the two spaces $\Bbb C^2$ in which these vectors live)
ordered in the following way: $j<k$ for $L$ even, while for $L$ odd, $j<k<\ell$ or
$\ell<j<k$ or $k<\ell<j$, $\ell$ being the label of the unmatched point.
In the case $L$ odd we also choose the
spin of the unmatched point
to be $\ket{+}_\ell$. The result is a vector in $({\Bbb C}^2)^{\otimes L}$ which satisfies
that its total $z$ spin $s^z=0$ or $s^z=1/2$ depending on parity.
For generic $q$ this mapping, which we call $S$, is injective.
In the case $L$ odd its image is in fact the whole $s^z=1/2$ subspace, of dimension ${L\choose(L-1)/2}$.
In the case $L$ even it is a subspace ($U_q({\goth sl}(2))$ singlets) of dimension 
${L!\over (L/2)!(L/2+1)!}$.
Furthermore, it is easy to show that the transfer matrices of the 
$O(1)$ loop model and of the six-vertex model are intertwined by $S$:
\eqn\equivspinloop{
\tilde{T}(t|z_1,\ldots,z_L)\, S =S\, T(t|z_1,\ldots,z_L) \ .
}
This is essentially a consequence of the fact that the $R$-matrix \Rspin\ involves just another
representation of the Temperley--Lieb algebra; explicitly, if $\check R=R{\cal P}$ where ${\cal P}$
switches factors of tensor products, then the formula \defcr\ holds with 
the Temperley--Lieb generator
$e= \pmatrix{0&0&0&0\cr 0&\omega&1&0\cr 0&1&\omega^{-1}&0\cr 0&0&0&0\cr}$.
The twist, necessary for $L$ even to take care of the arbitrariness in the ordering $j<k$ of
the indices, disappears for $L$ odd thanks to the unmatched point.
Henceforth, the loop model is equivalent to a sector of the six-vertex model.

At $q=-\e{i\pi/3}$, however, the mapping $S$ is no longer injective for $L$ odd.
In fact, one can show that the kernel of $S$ is exactly the same as the kernel of the 
projector $P_0$ of Sect.~2.3, so that the dimension of the image 
is nothing else than ${(2n+2)!\over (n+1)!(n+2)!}$, the dimension of the space of link
patterns of size $2n+2$. So at this special point, the {\it odd\/} six-vertex model is equivalent to
the {\it even}\/ loop model of size one more.

%%%%%%%%%%%%%%%%%%%%%%%%%%

\subsec{Mapping of odd size to even size}
\noindent
In analogy with the loop model, we now define a mapping of the six-vertex model
from $L=2n+1$ to $L=2n+2$. It is very simple: to a basis element $\alpha$,
that is a sequence of $2n+1$ spins, we
associate the new vector $\alpha-$ obtained by concatenating $\alpha$ and an extra minus spin.
Call $Q$ this mapping. We also need the projections $P_\pm$ within the model of size $2n+2$
which project onto the subspaces where the last spin is $\pm$, orthogonally to the subspace
where it is $\mp$, so that $P_+ + P_-=1$, $P_+ P_-=P_- P_+=0$.

We now have the two following properties of the $P_\pm$ and $Q$ operators:

(i) The subspace Im($P_+$) is stable under  $\tilde{T}(z_1,\ldots,z_{2n+1},\infty)$:
\eqn\stable{
P_+ \tilde{T}(t|z_1,\ldots,z_{2n+1},0)=\tilde{T}(t|z_1,\ldots,
z_{2n+1},0) P_+ \ .}
Indeed, if the last spectral parameter is zero, the corresponding matrix  $R_{2n+2,0}$ becomes
\eqn\Rzero{
R(0,t)=-q^{-1} \pmatrix{q^{-1}&0&0&0\cr 0&1&q^{-1}-q&0\cr 0&0& 1&0\cr 0&0&0&q^{-1}}
}
which means if the spin $2n+2$ is $+$ it will stay so.

(ii) At $q^3=1$, $Q$ intertwines the odd and even spin-chain transfer matrices, up to 
the projector $P_-$:
\eqn\oddevenspin{
Q \tilde{T}(t|z_1,\ldots,z_{2n+1})=P_- \tilde{T}(t|z_1,\ldots,z_{2n+1},0) Q\ .
}
This time we assume that the last spin is $-$. According to Eq.~\Rzero,
it might become $+$ after action
of $\tilde{T}$; so we project again with $P_-$. Now the action of $R_{2n+2,0}(0,t)$ becomes
simply the twist of the auxiliary space with its lower right submatrix, that is the
diagonal matrix $\left( {-q^{-1}\atop0}{0\atop -q^{-2}}\right)$: at $q^3=1$
this compensates exactly the twist $\Omega=\left( {-q\atop 0}{0\atop -q^{-1}}\right)$ and
we find that the r.h.s.\ of Eq.~\oddevenspin\ is simply the transfer matrix of odd size with
periodic boundary conditions and an additional spin at site $2n+2$ which is $-$.

\subsec{Polynomial eigenvector}
\noindent From now on we always set $q=-\e{i \pi/3}$, $\omega^{1/2}=\e{i\pi/6}$.
It is easy to check that $\tilde T$, just like $T$, possesses the eigenvalue $1$; we denote
by $\tilde\Psi$ (resp.\ $\tilde\Phi$) the corresponding polynomial eigenvector for $L$ odd (resp.\ even),
normalized so that its entries are coprime. 
This leaves an arbitrary constant in the normalization
which will be fixed below.

The various maps defined above allow us to write several relations between the various eigenvectors.
First and foremost, using properties (i) and (ii) of Sect.~4.3 we find that
\eqn\oddevenspinb{
Q \tilde\Psi(z_1,\ldots,z_{2n+1})=C P_- \tilde\Phi(z_1,\ldots,z_{2n+1},0)
}
or in components, $\tilde\Psi_\alpha(z_1,\ldots,z_{2n+1})=C\tilde\Phi_{\alpha-}(z_1,\ldots,z_{2n+1},0)$, with $C$ a normalization constant.
Let us prove this. In this paragraph parameters are omitted with the assumption that $z_{2n+2}=0$.
On the one hand, one can apply $\tilde\Psi$ to Eq.~\oddevenspin. We find:
$Q\tilde\Psi=P_- \tilde{T}Q\tilde\Psi$. On the other hand, decompose
$\tilde\Phi=P_+ \tilde\Phi+P_-\tilde\Phi$ and apply $P_-\tilde{T}$: using Eq.~\stable\ we find
$P_-\tilde{T}P_-\tilde\Phi=P_-\tilde\Phi$. Comparing, we find that $Q\tilde\Psi$
and $P_-\tilde\Phi$ must be proportional. The proportionality factor must be a constant
because components on both sides of the equation are coprime polynomials. 
Its numerical value will be determined below.

We also have the connection with loop models. For $L$ even this was already discussed in appendix B
of \DFZJ\ and we find simply
\eqn\evenspinloop{
S \Phi=\tilde\Phi
}
(up to a constant, which we fix to be 1).  However for $L$ odd, we only have
\eqn\oddspinloop{
S \Psi=B(z_1,\ldots,z_{2n+1}) \tilde{\Psi}
}
where $B$ is a polynomial to be determined in next section.
$B$ appears because $S$ is not injective (at $q^3=1$), so that the components of $S\Psi$
can have a non-trivial GCD. (Such a situation does not arise for $L$ even since $S$ is then injective).

%%%%%%%%%%%%%%%%%%%%%%%%%%%

\subsec{Sum rule}
\noindent
We now wish to compute the sum of entries of $\tilde\Psi$. 
In order to do so we can rely on the mapping to the even sized system:
\eqn\sumoddeven{
\sum_\alpha \tilde\Psi_\alpha(z_1,\ldots,z_{2n+1})=C
\sum_\alpha \tilde\Phi_{\alpha-}(z_1,\ldots,z_{2n+1},0)
}
and then use the mapping $S$ to the loop model. 
The normalization constant $C$  
will  be adjusted at the end of the computation. Via $S$,
each link pattern of size $2n+2$ contributes $\omega^{1/2}(\omega^{1/2}+\omega^{-1/2})^n$ 
to the r.h.s.\ of Eq.~\sumoddeven, so
$\sum_\alpha \tilde\Phi_{\alpha-}=\omega^{1/2}(\omega^{1/2}+\omega^{-1/2})^n \sum_\pi 
\Phi_\pi=\omega^{1/2}
3^{n/2} 3^{-n(n+1)/2}  % coming from the (q-q^{-1})
s_{\Y}(z_1,\ldots,z_{2n+2})$ 
according to \wellknown. So we have
\eqn\sumrule{
\sum_\alpha \tilde\Psi_\alpha(z_1,\ldots,z_{2n+1})=3^{-n^2/2}\omega^{1/2}
C s_{\Y}(z_1,\ldots,z_{2n+1})\ .}

To adjust the constant $C$ in \sumoddeven, we finally 
impose that the ``base component'' of $\tilde \Psi$, {\it i.e.} its 
smallest component in the homogeneous limit, be 1.
By \oddevenspinb, it is given by that of $\tilde\Phi$, itself 
proportional to that of $\Phi$,  
\eqn\basecomp{\tilde\Psi_{+\cdots+-\cdots-}= 
\omega^{(n+1)/2} \,\, C \!\!\!\!
\prod_{1\le i<j\le n+1\atop {\rm or}\ n+2\le i<j\le 2n+1}
\!\!\!\! {q\, z_i-q^{-1}z_j\over q-q^{-1}}
\ \prod_{i=n+2}^{2n+1} {q\, z_i\over q-q^{-1}}\ ,} 
 so at $z_i=1$,  $C^{-1} %\tilde\Psi_{+\cdots+-\cdots-}
=(-i\sqrt{3})^{-n} 
\omega^{(n+1)/2} q^n=3^{-n/2}\omega^{1/2}  $  
and the homogeneous sum rule is
\eqn\homosumrule{
{\sum_\alpha \tilde\Psi_\alpha} 
=3^{-n(n-1)/2} s_\Y(1^{2n+1})=3^{n/2}{\cal N}_n\ .}

As a side-product, by using $S$ directly on an odd-sized system, 
we can also compute the proportionality polynomial factor $B$:
\eqn\compB{
B \sum_\alpha \tilde\Psi_\alpha= (\omega^{1/2}+\omega^{-1/2})^n 
\sum_\pi \Psi_\pi
=3^{n/2}3^{-n^2} s_\Y s_{\pY}
}
so that $B=3^{-n(n-1)/2}\omega^{-1/2} C^{-1} s_\pY=3^{-n^2/2} s_\pY(z) $.

%%%%%%%%%%%%%%%%%%%%%%%%%

\subsec{Bilinear form and sum of squares}
\noindent
These sum rules possess obvious corollaries dealing with the sum of squares of components.
Indeed, for generic $q$ there is a bilinear form which to a pair of link patterns $\pi$ and $\pi'$ of $LP_{2n+2}$
associates $\braket{\pi}{\pi'}=(-q-1/q)^{\#}$ 
where $\#$ is the number of loops obtained by pasting together the two
diagrams along their common boundary. Via $S$, this bilinear form becomes simply diagonal in the
spin basis:
\eqn\bil{
\braket{\pi}{\pi'}=\sum_\alpha (S\pi)_\alpha (S\pi')_\alpha\ .
}
%(note that here we are in the subspace $s^z=0$ so the factor $q^{2s^z}$ which appears
%in general in the Markov trace is trivial).

We now specialize to $q=-\e{i\pi/3}$. Since $-q-1/q=1$, 
we have the identity $v\cdot\Phi=\braket{\Phi}{\pi_0}$ where $\pi_0$ is here any fixed link pattern.
More generally,
the bilinear form is degenerate of rank $1$
so that we can write (in size $L=2n+2$)
\eqn\degbil{
\sum_\alpha \tilde\Phi_\alpha^2 = \braket{\Phi}{\Phi}=\braket{\Phi}{\pi_0}\braket{\pi_0}{\Phi}=
(v\cdot\Phi)^2=3^{-n(n+1)}s_\Y^2
}
and in particular in the homogeneous limit, $\sum_\alpha \tilde\Phi_\alpha^2=A_{n+1}^2$, which is \sumspinev{b}.

As noticed in \RSunb, the even-size
twisted XXZ Hamiltonian is invariant under
simultaneous complex conjugation and spin reversal, and the ground state
$\tilde\Phi$ is chosen to be invariant under this operation.
Thus $\tilde\Phi_{\alpha+}=\overline{\tilde\Phi_{\overline{\alpha}-}}$, and
\degbil\ may be rewritten as
$$ \braket{\Phi}{\Phi}= 
(v\cdot\Phi)^2=3^{-n(n+1)}s_\Y^2= \sum_\alpha \tilde\Phi_\alpha^2 = 
\sum_{\alpha'} (\tilde\Phi_{\alpha'-}^2+\tilde\Phi_{\alpha'+}^2)=
2 \Re\Big( \sum_{\alpha'} \tilde\Phi_{\alpha'-}^2\Big)\ .$$
Using Eq.~\oddevenspinb, we can then derive the sum rule for the squares
of the $\tilde\Psi$, in odd size. 
By taking the real part of
$C^{-2}\sum_{\alpha'} \tilde\Psi_{\alpha'}^2(z)=\sum_{\alpha'} \tilde\Phi_{\alpha'-}^2(z,0)$, (the $\tilde \Psi$ are real), 
and using the value of $C$ determined above, we get 
\eqn\realpt{
\sum_{\alpha'} \tilde\Psi_{\alpha'}^2={\Re \sum_{\alpha'} \tilde\Phi_{\alpha'-}^2
\over \Re C^{-2}}= 3^{-n^2}s_\Y^2(z_1,\ldots,z_{2n+1},0)\ .}
In the homogeneous limit, we get 
\eqn\sumrulqu{\sum_{\alpha'} \tilde\Psi_{\alpha'}^2= 
\CN_n^2=A_{{\rm HT}}(2n+1)}
as announced in  \normsq. 

In contrast, the alledged sum rule \sumspinev{c} 
seems to be of a different nature,
since its deformation by spectral parameters does not involve symmetric
functions of $z_1,\ldots, z_{2n}$. % <<<

\newsec{Conclusion}
\noindent
In this paper, we have proved  multi-parameter sum rules for the 
components of the groundstate vector of the inhomogeneous $O(1)$ 
loop model on a punctured disk of odd or even perimeter, 
or of its XXZ chain counterpart. 
Our strategy has relied on the construction of projection 
operators onto components of the 
groundstate vector of the same model, but on 
a non-punctured disk of perimeter one more,
in which the extra inhomogeneity (spectral parameter) is 
taken to either $0$ or $\infty$.
The new ingredient here is the use of knot-theoretic 
crossing operators to relate
the non-crossing link patterns of both models.
Admittedly, our proof in the odd case still relies on a technical assumption 
that we have been unable to prove in general, namely
the fact that the two 
Schur functions $s_\Y$ and $s_\pY$ of Sect.~2.5 have no common factor.
It would be highly desirable to fill this gap. 

\medskip

In the derivations of Sect.~2 and 3 we made an assumption 
of minimal degree, {\it viz}\/ that the base component of $\Psi$ or
$\Phi$ is given by \tobe, \fundaeven. The justification of this assumption
could be done as in \DFZJ, through the use of the Bethe Ansatz, 
or by appealing to the representation theory of the $q$KZ equation. 
\medskip

As already mentionned in Sect.~1, our formulae for the 
sums of components of the $O(1)$ loop model lead to the
same expressions as those derived by Razumov and Stroganov
 for the partition 
functions of the square-ice model with adequate boundary conditions. 
The latter are designed so as to make 
a bijection between the states of the lattice model and  
half turn symmetric ASMs of size $L\times L$.  
In  \RSqu, the partition function of the
six-vertex model on a $L\times \lfloor \oh (L+1)\rfloor$ grid
with such ``half-turn symmetric'' boundary conditions \Ku\
was computed  as a function of spectral parameters $x_i$   
and $y_j$, $i,j=1,\ldots,\lfloor{L+1\over 2}\rfloor $.  
It is a symmetric function of the $x$'s and
the $y$'s, thanks to the Yang--Baxter equation, 
and it  was shown to be a product of
two factors. For the special value of the crossing parameter 
corresponding to our $q=-\e{i\pi/3}$ (their $a=-q$), 
and in the odd size case ($L=2n+1$), upon specialization of the last 
spectral parameter $ y_{n+1}=x_{n+1}$, 
it becomes a completely 
 symmetric function of all its spectral parameters and it
 reads (in our notations) 
\eqnn\pnfhRS
$$\eqalignno{
\!\!\!\!\!\!\!
Z_{{\rm HT}}&(x_1,\ldots,x_n, y_1,\ldots,y_n)
&L=2n\cr
&= (-3)^n  \prod_{i=1}^{n}(x_i y_i)^{1-2n}
s_{\ppY}(x_1^2,\ldots, y_n^2) s_{\pY}(x_1^2,\ldots, y_n^2)\cr
Z_{{\rm HT}}&(x_1,\ldots,x_{n+1}, y_1,\ldots,y_n, y_{n+1}=x_{n+1})
&L=2n+1\cr
&= 3^n \prod_{i=1}^{n+1}x_i^{-2n}\prod_{i=1}^{n}y_i^{-2n}
 s_{\Y}(x_1^2,\ldots,x^2_{n+1},y^2_1,\ldots, y_n^2) 
s_{\pY}(x_1^2,\cdots, y_n^2) 
&\pnfhRS\cr
}
$$
that is, up to a trivial factor ($(-3)^{n^2}\prod (x_j y_j)^{1-2n}$, 
  resp. $3^{n(n+1)}\prod x_j^{-2n}\prod y_j^{-2n}$),  
the same expression 
as $\sum_\pi \Phi^*$, resp. $\sum_\pi \Psi$.  
Recall that the partition function, upon some specialization of its spectral
parameters, yields the ``refined $x$-enumerations'' of HTSASMs. The agreement
between $Z_{{\rm HT}}(z)$ and the sum of the components
is a further indication of the detailed 
connection between the two sets of problems: the determination of the
Perron--Frobenius  eigenvectors of the $O(1)$ loop models and 
the  counting of HTSASMs. More precisely it gives support to the 
strong RS conjecture that $\Psi_\pi$ or $\Phi^*_\pi$ count the number
of HTSFPL configurations of link pattern $\pi$ (see Sect. 1). 

At this stage, our results do not, however, provide any 
direct way to verify the conjectures
 about the largest components of the Perron--Frobenius 
eigenvectors (see Sect. 1), nor to even confirm the empirical
observation that in all cases (but the twisted $\tilde \Phi$)
all components are positive integers once the smallest one
is normalized to 1.

\medskip

Another interesting question is what survives of these constructions for generic values of $q$
(not at the RS point). One expects in particular from previous experience 
\refs{\DFZJc,\DFZJd}  
that the point $q=-1$ have some geometrical interpretation, involving (multi-)degrees of some
matrix schemes, and we indeed checked for odd sizes that all components of $\Psi$ become
(possibly vanishing) non-negative integers in the homogeneous limit $z_i\to 1$.
It is not too difficult to convince oneself that 
the correct definition of the projection operators $P_0$ and $P_\infty$ of Sect.~2 
 requires switching to different conventions
for the crossing operators $T_i\equiv \Rc_{i,i+1}(z,0)$. Via these projections, we may relate the vector 
$\Psi$ at generic $q$ to solutions of a modified $q$KZ equation for size one more, but with the extra
spectral parameter sent to $0$ or $\infty$. In the case of size $2n+1$, we have found that $P_0$ creates
a solution of this modified $q$KZ equation in size $2n+2$ with $z_{2n+2}\to 0$, whose base
component $\Theta_{\pi_0}$, corresponding to the fully nested link pattern $\pi_0$
relating $i$ to $2n+3-i$, $i=1,2,\ldots,n+1$, simply reads:
\eqn\compofond{ \Theta_{\pi_0}(z_1,\ldots,z_{2n+1})= \Phi_{\pi_0}(z_1,\ldots,z_{2n+1},0)\ 
G(z_1,\ldots,z_{2n+1},\infty) }
where $\Phi_{\pi_0}$ stands as before for the (completely factorized) base
component of the loop model on an unpunctured disk of size $2n+2$, while $G$ stands
for the so-called Gaudin determinant
\eqn\gaudin{\eqalign{ G(z_1,\ldots,z_{2n+2})&= 
{\prod_{1\leq i,j \leq n+1}(qz_i-q^{-1}z_{n+j+1})(q^{1/2}z_i -q^{-1/2}z_{n+j+1})\over
\prod_{1\leq i<j \leq n+1}(z_i-z_{j})(z_{n+i+1}-z_{n+j+1})} \cr
&\times \det_{1\leq i,j\leq n+1} \left({1\over qz_i-q^{-1}z_{n+j+1}}
{1\over q^{1/2}z_i -q^{-1/2}z_{n+j+1}}\right) \cr}}
and the symbol $\infty$ in place of $z_{2n+2}$ stands as usual for the suitably normalized 
large $z_{2n+2}\to \infty$ limit. Note that $P_\infty$ does just the opposite, namely
interchanges $0$ and $\infty$ in Eq.~\compofond.
This is remarkably reminiscent of the higher degree solution to the $q$KZ equation found by Pasquier
\Pas\ on the even size unpunctured disks, for which the cyclicity condition analogous to that of 
the second line of eq.\compo\ has the {\it same}\/ shift $s=q^3$ (as opposed to $s=q^6$ for the minimal 
degree solution).

\centerline{\bf Acknowledgments}

We thank M. Kasatani and V. Pasquier, who are currently working
on related issues, for interesting discussions.
We acknowledge the support of the Geocomp project 
(ACI Masse de Donn\'ees), the  
European networks ``ENIGMA", grant MRT-CT-2004-5652 and ``ENRAGE", 
grant MRTN-CT-2004-5616, and
the ANR program ``GIMP'', ANR-05-BLAN-0029-01.

\listrefs
\end